\documentstyle[12pt]{article}
\newlength{\extraspace}
\newlength{\extraspaces}
\newcommand{\ba}{\begin{eqnarray}
\addtolength{\abovedisplayskip}{\extraspaces}
\addtolength{\belowdisplayskip}{\extraspaces}
\addtolength{\abovedisplayshortskip}{\extraspace}
\addtolength{\belowdisplayshortskip}{\extraspace}}
\newcommand{\ea}{\end{eqnarray}}

\newcommand{\nonu}{\nonumber \\[.5mm]}
\newcommand{\A}{&\!\!\!}

\begin{document}

\thispagestyle{empty}

\hfill \parbox{3.5cm}{hep-th/0101178 \\ SIT-LP-01/01}
\vspace*{1cm}
\begin{center}
{\bf  NEW SUPERSYMMETRY ALGEBRA ON GRAVITATIONAL   \\
INTERACTION OF NAMBU-GOLDSTONE FERMION } \\[20mm]
{Kazunari SHIMA and Motomu TSUDA} \\[2mm]
{\em Laboratory of Physics, Saitama Institute of Technology}
\footnote{e-mail: shima@sit.ac.jp, tsuda@sit.ac.jp}\\
{\em Okabe-machi, Saitama 369-0293, Japan}\\[2mm]
{January 2001}\\[15mm]


\begin{abstract}
The spacetime symmetries of SGM action proposed as the gravitational 
coupling of N-G fermions are investigated.
The commutators of new nonlinear supersymmetry (NL SUSY) transformations 
form a closed algebra, which reveals N-G fermion (NL SUSY) nature 
and a generalized general coordinate transformation. 
A generalized local Lorentz transformation, 
which forms a closed algebra, is also introduced.

\end{abstract}
\end{center}

\newpage
The supersymmetry (SUSY)\cite{wz} and its spontaneous 
breakedown are the essential notions to 
unify spacetime and matter. 
It is well understood that the Nambu-Goldstone (N-G) 
fermion with spin 1/2 would appear 
in the spontaneous breakdown of SUSY 
and that it can be converted to the longitudunal 
components of the spin 3/2 field (gravitino) 
through the superHiggs mechanism. 
This is demonstrated explicitly by the introduction 
of the local gauge coupling of Volkov-Akulov (V-A) 
model\cite{va} of a nonlinear realization of SUSY (NL SUSY) 
to the supergravity (SUGRA) gauge multiplet\cite{DZ}. 

In the previous papers \cite{ks1} and \cite{ks3}, 
a supersymmetric composite unified model for spacetime 
and matter, superon-graviton model (SGM) based upon SO(10) 
super-Poincar\'e algebra, is proposed. 
The fundamental action which is the analogue of Einstein-Hilbert(E-H) 
action of general relativity(GR) describes 
the gravitational interaction of the spin 1/2 N-G fermions 
of V-A NL SUSY\cite{va} regarded as the fundamental objects 
(superon-quintet) for matter. 
SGM may be the most economical model that accomodates 
all observed particles in a single irreducible 
representation of a (semi)simple group. 
The NL SUSY may give a framework to describe 
the unity of nature from the compositeness viewpoint for matter.  
In SGM all particles participating in (super)Higgs mechanism 
except graviton are composites of N-G fermions, superons. 
In ref.\cite{ks3}, we have constructed the gauge invariant 
SGM action based upon a new NL SUSY transformation 
and clarified the systematics in the unified model building. 

We have further extended the framework \cite{ks3} 
to N-G fermion with the higher spin in ref.\cite{st1} 
and obtained a new action(spin 3/2 SGM) analogous to 
E-H action of GR,  
where the action of spin 3/2 
N-G fermion written down by Baaklini 
as a nonlinear realization of a superalgebra 
containing a vector-spinor generator\cite{rikuzw} 
is extended systematically to the curved spacetime. 

In this letter we demonstrate the (spacetime) symmetry of 
our SGM actions, especially we compute the commutators 
of new NL SUSY transformations on gravitational 
interaction of N-G fermion with spin 1/2 and 3/2 
introduced in refs.\cite{ks3} and \cite{st1}. 
We will show that the commutators of spacetime symmetries form 
a closed algebra, which reveals N-G  (NL SUSY) 
nature of fermions and the invariances at least 
under a generalized general coordinate 
and a generalized local Lorentz transformations. 
\vspace{.7cm} \\
{\it [I] The gravitational interaction of spin 1/2 N-G fermion:} 
\vspace{.2cm} \\
We first summarize new NL SUSY transformations 
introduced in ref.\cite{ks3}. 
In the arguments of SGM\cite{ks3}, 
we have regarded that the coset space (N-G fermion) SL(2,C) coordinates 
in addition to the SO(3,1) Lorentz coordinates are embedded at every 
curved spacetime point and introduce formally 
a new vierbein field $w{^a}_{\mu}(x)$ 
through the NL SUSY invariant differential 
forms $\omega^a$ of V-A\cite{va} as follows: 
\footnote{
Latin $(a,b,..)$ and Greek $(\mu,\nu,..)$ are the indices 
for local Lorentz and general coordinates, respectively.} 
\ba
\A \A \omega^a = w{^a}_{\mu} dx^{\mu}, 
\label{om} \\
\A \A w{^a}_{\mu}(x) = e{^a}_{\mu}(x) + t{^a}_{\mu}(x), 
\label{new-w}
\ea
where $e{^a}_{\mu}(x)$ is the vierbein of Einstein 
genaral relativity theory (EGRT) and $t{^a}_{\mu}(x)$ 
is defined by 
\footnote{
In our convention ${1 \over 2}\{ \gamma^a, \gamma^b \} = \eta^{ab} 
= (+, -, -, -)$ 
and $\sigma^{ab} = {i \over 4}[\gamma^a, \gamma^b]$.} 
\begin{equation}
t{^a}_{\mu}(x) = ia \bar{\psi} \gamma^a 
\partial_{\mu} \psi 
\label{T1/2}
\end{equation} 
for spin 1/2 Majorana N-G field $\psi$. 
In Eq.(\ref{T1/2}) $a$ is an arbitrary constant 
with the dimension of the fourth power of length 
(i.e., a fundamental volume of spacetime). 
We can easily obtain the inverse of the new vierbein, 
$w{_a}^{\mu}$, in the power series of $t{^a}_{\mu}$ 
which terminates with $t^4$: 
\begin{equation}
w{_a}^{\mu} = e{_a}^{\mu} 
- t{^{\mu}}_a + t{^{\rho}}_a t{^{\mu}}_{\rho} - \dots . 
\end{equation}
Similarly a new metric tensor $s^{\mu\nu}(x)$ 
is formally introduced in the abovementioned 
curved spacetime as follows: 
\begin{equation}
s^{\mu\nu}(x) \equiv w{_a}^{\mu}(x) w^{a \nu}(x). 
\end{equation}
It is straightforward to show 
${w_a}^{\mu} w_{b \mu} = \eta_{ab}$, 
$s_{\mu \nu}{w_a}^{\mu} {w_b}^{\mu} = \eta_{ab}$, ..etc. 

In order to obtain simply the action 
in the abovementioned curved spacetime, 
which is invariant at least under GL(4,R), NL SUSY 
and local Lorentz transformations, 
we follow formally EGRT as performed in SGM\cite{ks3}. 
That is, we require that the (mimic) vierbein 
$w{^a}_{\mu}(x)$ and the metric $s^{\mu\nu}(x)$ 
should have formally a general coordinate transformation 
under the supertranslations: 
\begin{equation}
\delta x^{\mu} = - \xi^{\mu}, \quad 
\delta \psi = \zeta, 
\label{st1/2}
\end{equation}
where $\xi^{\mu} = ia \bar{\zeta} \gamma^{\mu} {\psi}$ and 
$\zeta$ is a constant spinor parameter.  \\
Remarkably the following nonlinear new (super)transformations 
\ba
\A \A \delta \psi(x) 
      = \zeta + ia (\bar{\zeta} \gamma^{\mu} {\psi}) 
      \partial_{\mu} \psi, 
\label{nst1/2-p} \\
\A \A \delta e{^a}_{\mu}(x) 
      = ia (\bar{\zeta} \gamma^{\rho} {\psi}) 
      D_{[\rho} e{^a}_{\mu]}, 
\label{nst1/2-e}
\ea
induce the desirable transformations on $w{^a}_{\mu}(x)$ 
and $s^{\mu\nu}(x)$ as follows: 
\footnote{
Throughout the paper $D_{\mu}$ is the covariant derivative of GL(4,R) with the symmetric affine connection.} 
\ba
\A \A \delta_{\zeta} w{^a}_{\mu} 
      = \xi^{\nu} \partial_{\nu} w{^a}_{\mu} 
      + \partial_{\mu} \xi^{\nu} w{^a}_{\nu}, 
\label{gc-w} \\
\A \A \delta_{\zeta} s_{\mu\nu} = \xi^{\kappa} 
      \partial_{\kappa} s_{\mu\nu} + \partial_{\mu} 
      \xi^{\kappa} s_{\kappa\nu} 
      + \partial_{\nu} \xi^{\kappa} s_{\mu\kappa}. 
\label{gc-s}
\ea
That is, $w{^a}_{\mu}(x)$ and $s^{\mu\nu}(x)$ 
have general coordinate transformations under 
the new supertransformations (\ref{nst1/2-p}) 
and (\ref{nst1/2-e}). 

In addition, to embed simply the local Lorentz invariance 
we follow EGRT formally and require that the new vierbein 
$w{^a}_{\mu}(x)$ should also have formally 
a local Lorentz transformation, i.e., 
\begin{equation}
\delta_L w{^a}_{\mu} 
= \epsilon{^a}_b w{^b}_{\mu} 
\label{Lrw}
\end{equation} 
with the local Lorentz transformation parameter 
$\epsilon_{ab}(x) = (1/2) \epsilon_{[ab]}(x)$.   \\
Interestingly,  we find that the following (generalized) 
local Lorentz transformations on  $\psi$ and $e{^a}_{\mu}$
\ba
\A \A \delta_L \psi(x) = - {i \over 2} \epsilon_{ab} 
      \sigma^{ab} \psi, 
\label{Lr1/2-p} \\
\A \A \delta_L {e^{a}}_{\mu}(x) = \epsilon{^a}_b e{^b}_{\mu} 
      + {a \over 4} \varepsilon^{abcd} 
      \bar{\psi} \gamma_5 \gamma_d \psi 
      (\partial_{\mu} \epsilon_{bc}) 
\label{Lr1/2-e}
\ea
induce the desirable transformation (\ref{Lrw}). 
[ Note that the equation (\ref{Lr1/2-e}) reduces to the familiar 
form of the Lorentz transformations 
if the global transformations are considered, e.g., $\delta_{L}g_{\mu\nu}=0$.] 

Therefore, replacing $e{^a}_{\mu}(x)$ 
in E-H Lagrangian of GR 
by the new vierbein $w{^a}_{\mu}(x)$ 
we obtain the analogue of E-H Lagrangian 
which is invariant under (\ref{nst1/2-p}), (\ref{nst1/2-e}), 
(\ref{Lr1/2-p}) and (\ref{Lr1/2-e})\cite{ks3}: 
\ba
\A \A L = - {c^3 \over 16{\pi}G} 
          \vert w \vert(\Omega + \Lambda ), 
\label{L-SGM} \\
\A \A \vert w \vert = {\rm det} w{^a}_{\mu} 
      = {\rm det}(e{^a}_{\mu} + t{^a}_{\mu}), 
\ea
where the overall factor is now fixed uniquely to 
${-c^3 \over 16{\pi}G}$, 
$e{_a}^{\mu}(x)$ is the vierbein of EGRT and 
$\Lambda$ is a probable cosmological constant. 
$\Omega$ is a mimic new scalar curvature 
analogous to the Ricci scalar curvature $R$ of EGRT. 
The explicit expression of $\Omega$ is obtained 
by just replacing $e{_a}^{\mu}(x)$ in Ricci scalar $R$ 
of EGRT by $w{_a}^{\mu}(x) = e{_a}^{\mu} + t{_a}^{\mu}$, 
which gives the gravitational interaction of 
$\psi(x)$. 
The lowest order term of $a$ in the action (\ref{L-SGM}) 
gives the E-H action of GR. 
And in the flat spacetime, i.e., 
$e{_a}^{\mu}(x) \rightarrow \delta{_a}^{\mu}$, 
it reduces to V-A model\cite{va} 
with ${\kappa}^{-1} = {c^3 \over 16{\pi}G}{\Lambda}$. 
Therefore our model(SGM) needs 
a non-zero (small) cosmological constant. 
Also from the low energy theorem viewpoint, 
the coupling constant of N-G fermion(superon) to 
the vacuum via the supercurrent is given 
$({c^3 \over 16{\pi}G}{\Lambda})^{1 \over 2}$.

The commutators of two new supersymmetry transformations 
(\ref{nst1/2-p}) and (\ref{nst1/2-e}) 
on $\psi(x)$ and $e{_a}^{\mu}(x)$ 
are calculated straightforwardly as \cite{ks3} 
\ba
\A \A [\delta_{\zeta_1}, \delta_{\zeta_2}] \psi 
      = \{ 2ia (\bar{\zeta}_2 \gamma^{\mu} \zeta_1) 
      - \xi_1^{\rho} \xi_2^{\sigma} e{_a}^{\mu} 
      (D_{[\rho} e{^a}_{\sigma]}) \} 
      \partial_{\mu} \psi, \\
\A \A [\delta_{\zeta_1}, \delta_{\zeta_2}] e{^a}_{\mu} 
      = \{ 2ia (\bar{\zeta}_2 \gamma^{\mu} \zeta_1) 
      - \xi_1^{\sigma} \xi_2^{\lambda} e{_c}^{\rho} 
      (D_{[\sigma} e{^c}_{\lambda]}) \} 
      D_{[\rho} e{^a}_{\mu]} 
      - \partial_{\mu} (\xi_1^{\rho} \xi_2^{\sigma} 
      D_{[\rho} e{^a}_{\sigma]}). 
\ea
These can be rewritten in the following familiar forms 
of the general coordinate transformations 
\ba
\A \A [\delta_{\zeta_1}, \delta_{\zeta_2}] \psi 
      = \Xi^{\mu} \partial_{\mu} \psi, 
\label{com1/2-p} \\
\A \A [\delta_{\zeta_1}, \delta_{\zeta_2}] e{^a}_{\mu} 
      = \Xi^{\rho} \partial_{\rho} e{^a}_{\mu} 
      + e{^a}_{\rho} \partial_{\mu} \Xi^{\rho}, 
\label{com1/2-e}
\ea
where $\Xi^{\mu}$ is defined by 
\begin{equation}
\Xi^{\mu} = 2ia (\bar{\zeta}_2 \gamma^{\mu} \zeta_1) 
      - \xi_1^{\rho} \xi_2^{\sigma} e{_a}^{\mu} 
      (D_{[\rho} e{^a}_{\sigma]}). 
\end{equation}
Therefore, the equations (\ref{nst1/2-p}), (\ref{nst1/2-e}), 
(\ref{com1/2-p}) and (\ref{com1/2-e}) 
reveal N-G fermion (NL SUSY) nature of $\psi(x)$, 
non-N-G nature of $e{_a}^{\mu}(x)$ corresponding to SGM scenario 
and  generalized general coordinate transformations, 
which form a closed algebra. 

Similarly, it is interesting to compute the commutator of 
the local Lorentz transformation on $e{_a}^{\mu}(x)$ 
of Eq.(\ref{Lr1/2-e}). It is calculated as 
\begin{equation}
[\delta_{L_{1}}, \delta_{L_{2}}] e{^a}_{\mu} 
= \beta{^a}_b e{^b}_{\mu} 
+ {a \over 4} \varepsilon^{abcd} \bar{\psi} 
\gamma_5 \gamma_d \psi 
(\partial_{\mu} \beta_{bc}), 
\label{comLr1/2}
\end{equation}
where $\beta_{ab}=-\beta_{ba}$ is defined by 
\begin{equation}
\beta_{ab} = \epsilon_{2ac}\epsilon{_1}{^c}_{b} -  \epsilon_{2bc}\epsilon{_1}{^c}_{a}.
\label{Lpara}
\end{equation}
Remarkably, the equations (\ref{Lr1/2-e}) and (\ref{comLr1/2}) 
explicitly reveal a generalized local Lorentz transformation 
with the parameters $\epsilon_{ab}$ and $\beta_{ab}$ respectively, 
which shows the closure of the algebra. 
As for the internal symmetry, the global SO(N) symmetry can be introduced 
by replacing $\psi(x) \rightarrow  \psi^{i}(x),(i=1,2, \dots, N)$.  \\
These arguments show that our  action (\ref{L-SGM}) with (\ref{T1/2})                                              
is invariant at least under 
\begin{equation}
[{\rm global\ NL\ SUSY}] \otimes [{\rm local\ GL(4,R)}] \otimes [{\rm local\ Lorentz}] \otimes [{\rm global\ SO(N)}].   
\end{equation}
SGM \cite{ks3} for spacetime and matter is the case with N=10.
\vspace{.7cm} \\
{\it [II] The gravitational interaction of spin 3/2 N-G fermion:} 
\vspace{.2cm} \\
The arguments are completely parallel with the spin 1/2 case.
For the extension of the framework \cite{ks3} 
to N-G fermion with spin 3/2\cite{st1}, 
a new vierbein field ${w^{a}}_{\mu}(x) 
(= e{^a}_{\mu}(x) + t{^a}_{\mu}(x))$ 
is also introduced through the NL SUSY invariant 
differential forms $\omega_{a}$ of Baaklini\cite{rikuzw} 
as (\ref{om}) and (\ref{new-w}) with 
\begin{equation}
t{^a}_{\mu}(x)= ia \varepsilon^{abcd} \bar{\psi}_b 
\gamma_c \gamma_5 \partial_{\mu} \psi_d 
\label{T3/2}
\end{equation} 
for spin 3/2 Majorana N-G field $\psi_{a}(x)$. 
As in the case for gravitational interaction of 
spin 1/2 N-G fermion, we require that the (mimic) vierbein 
$w{^a}_{\mu}(x)$ and the metric $s^{\mu\nu}(x) \equiv w^{a\mu}(x) w{_a}^{\mu}(x)$ 
should have formally a general coordinate transformation 
under the supertranslations: 
\begin{equation}
\delta x^{\mu} = - \xi^{\mu}, \quad 
\delta \psi^a = \zeta^a, 
\label{st3/2}
\end{equation}
where $\xi^{\mu} = ia \varepsilon^{\mu\nu\rho\sigma} 
\bar{\psi}_{\nu} \gamma_{\rho} \gamma_5 \zeta_{\sigma}$ 
and $\zeta^{a}$ is a constant Majorana spinor parameter  with spin 3/2. 
Remarkably we find that the following nonlinear new  supertransformations 
\ba
\A \A \delta \psi^a(x) 
      = \zeta^a - ia (\varepsilon^{\mu\nu\rho\sigma} 
      \bar{\psi}_{\nu} \gamma_{\rho} \gamma_5 
      \zeta_{\sigma}) \partial_{\mu} \psi^a, 
\label{nst3/2-p} \\
\A \A \delta e{^a}_{\mu}(x) 
      = - ia \varepsilon^{\rho\nu\sigma\lambda} 
      \bar{\psi}_{\nu} \gamma_{\sigma} \gamma_5 
      \zeta_{\lambda}D_{[\rho} e{^a}_{\mu]} 
\label{nst3/2-e}
\ea
induce the desirable transformations on $w{^a}_{\mu}(x)$ 
and $s^{\mu\nu}(x)$ as (\ref{gc-w}) and (\ref{gc-s}). 
That is, $w{^a}_{\mu}(x)$ and $s^{\mu\nu}(x)$ 
have general coordinate transformations under 
the new supertransformations (\ref{nst3/2-p}) 
and (\ref{nst3/2-e}). 

As for the Lorentz invariance we again require that the new vierbein 
$w{^{a}}_{\mu}(x)$ should have formally 
a local Lorentz transformation (\ref{Lrw}). 
Then we find that the following (generalized) 
local Lorentz transformations 
\ba
\A \A \delta_L \psi^a(x) = \epsilon{^a}_b \psi^b 
      - {i \over 2} \epsilon_{bc} \sigma^{bc} \psi^a, 
\label{Lr3/2-p} \\
\A \A \delta_L e{^a}_{\mu}(x) = \epsilon{^a}_b e{^b}_{\mu} 
      - ia \varepsilon^{abcd} 
      \{ \bar{\psi}_b \gamma_c \gamma_5 \psi_e 
      (\partial_{\mu} \epsilon{_d}^e) 
      - {i \over 4} \varepsilon{_c}^{efg} 
      \bar{\psi}_b \gamma_g \psi_d 
      (\partial_{\mu} \epsilon_{ef}) \} 
\label{Lr3/2-e}
\ea
induce the desirable transformation (\ref{Lrw}). 
[The equation (\ref{Lr3/2-e}) also reduces 
to the familiar form of the Lorentz transformations 
if the global transformations are considered.] 

Therefore, as in spin 1/2 SGM case,  replacing $e{^a}_{\mu}(x)$ 
in E-H Lagrangian of GR 
by the new vierbein $w{^a}_{\mu}(x)$ defined by (\ref{new-w}) 
with (\ref{T3/2}), 
we obtain the Lagrangian of the same form as (\ref{L-SGM}), 
which is invariant under (\ref{nst3/2-p}), (\ref{nst3/2-e}), 
(\ref{Lr3/2-p}) and (\ref{Lr3/2-e}). 

The commutators of two new supersymmetry transformations 
(\ref{nst3/2-p}) and (\ref{nst3/2-e}) 
on $\psi^{a}(x)$ and  $e{_a}^{\mu}(x)$ 
are now calculated as\cite{st1} 
\ba
\A \A [\delta_{\zeta_1}, \delta_{\zeta_2}] \psi^{a} 
      = \{ 2ia (\varepsilon^{\mu bcd} \bar{\zeta}_{2b} 
      \gamma_c \gamma_5 \zeta_{1d}) 
      - \xi_1^{\rho} \xi_2^{\sigma} e{_a}^{\mu} 
      (D_{[\rho} e{^a}_{\sigma]}) \} 
      \partial_{\mu} \psi^a, \\
\A \A [\delta_{\zeta_1}, \delta_{\zeta_2}] e{^a}_{\mu} 
      = \{ 2ia (\varepsilon^{\rho bcd} \bar{\zeta}_{2b} 
      \gamma_c \gamma_5 \zeta_{1d}) 
      - \xi_1^{\sigma} \xi_2^{\lambda} e{_c}^{\rho} 
      (D_{[\sigma} e{^c}_{\lambda]}) \} 
      D_{[\rho} e{^a}_{\mu]} \nonu
\A \A \hspace{2.5cm} 
      - \partial_{\mu} (\xi_1^{\rho} \xi_2^{\sigma} 
      D_{[\rho} e{^a}_{\sigma]}). 
\ea
These can be rewritten as 
\ba
\A \A [\delta_{\zeta_1}, \delta_{\zeta_2}] \psi^{a} 
      = \Xi^{\mu} \partial_{\mu} \psi^{a}, 
\label{com3/2-p} \\
\A \A [\delta_{\zeta_1}, \delta_{\zeta_2}] e{^a}_{\mu} 
      = \Xi^{\rho} \partial_{\rho} e{^a}_{\mu} 
      + e{^a}_{\rho} \partial_{\mu} \Xi^{\rho}, 
\label{com3/2-e}
\ea
where $\Xi^{\mu}$ is now a generalized gauge parameter defined by 
\begin{equation}
\Xi^{\mu} = 2ia (\varepsilon^{\mu bcd} \bar{\zeta}_{2b} 
      \gamma_c \gamma_5 \zeta_{1d}) 
      - \xi_1^{\rho} \xi_2^{\sigma} e{_a}^{\mu} 
      (D_{[\rho} e{^a}_{\sigma]}). 
\end{equation}
Therefore the equations (\ref{nst3/2-p}), (\ref{nst3/2-e}), 
(\ref{com3/2-p}) and (\ref{com3/2-e}) 
reveal N-G fermion (NL SUSY) nature of $\psi^{a}(x)$, 
non-N-G nature of $e{_a}^{\mu}(x)$ 
and a generalized general coordinate transformation, 
which form a closed algebra. 

Also, the commutator of 
the local Lorentz transformation on $e{_a}^{\mu}(x)$ 
of Eq.(\ref{Lr3/2-e}) is calculated as 
\begin{equation}
[\delta_{L_{1}}, \delta_{L_{2}}] e{^a}_{\mu} 
= \beta{^a}_b e{^b}_{\mu} 
- ia \varepsilon^{abcd} 
\{ \bar{\psi}_b \gamma_c \gamma_5 \psi_e 
(\partial_{\mu} \beta{_d}^e) 
- {i \over 4} \varepsilon{_c}^{efg} 
\bar{\psi}_b \gamma_g \psi_d 
(\partial_{\mu} \beta_{ef}) \} 
\label{comLr3/2}
\end{equation}
where $\beta_{ab}$ is the same as (\ref{Lpara}). 
The equations (\ref{Lr3/2-e}) and (\ref{comLr3/2}) 
explicitly reveal a generalized local Lorentz transformation 
with the parameters $\epsilon_{ab}$ and $\beta_{ab}$, 
which forms a closed algebra.     \\
Therefore our action (\ref{L-SGM}) with (\ref{T3/2}), which is the analogue of 
the E-H action of GR, is invariant at least under   
\begin{equation}
[{\rm global\ NL\ SUSY}] \otimes [{\rm local\ GL(4,R)}] 
\otimes [{\rm local\ Lorentz}] \otimes [{\rm global\ SO(N)}],
\end{equation}
when it is extended to global SO(N). 
          
SGM formalism \cite{ks3}, i.e. the action (\ref{L-SGM}) can be  generalized 
to the spacetime with extra dimensions and to the inclusion of the non-abelian internal symmetries. 
It may give a potential new framework for the simple unification of spacetime and matter.

\vskip 30mm

One of the authors(K.S.)  would like to thank J. Wess for his useful suggestions on the algebra. 
The work of M. Tsuda is supported in part by  High-Tech research program of
Saitama Institute of Technology.

\newpage

%
\newcommand{\NP}[1]{{\it Nucl.\ Phys.\ }{\bf #1}}
\newcommand{\PL}[1]{{\it Phys.\ Lett.\ }{\bf #1}}
\newcommand{\CMP}[1]{{\it Commun.\ Math.\ Phys.\ }{\bf #1}}
\newcommand{\MPL}[1]{{\it Mod.\ Phys.\ Lett.\ }{\bf #1}}
\newcommand{\IJMP}[1]{{\it Int.\ J. Mod.\ Phys.\ }{\bf #1}}
\newcommand{\PR}[1]{{\it Phys.\ Rev.\ }{\bf #1}}
\newcommand{\PRL}[1]{{\it Phys.\ Rev.\ Lett.\ }{\bf #1}}
\newcommand{\PTP}[1]{{\it Prog.\ Theor.\ Phys.\ }{\bf #1}}
\newcommand{\PTPS}[1]{{\it Prog.\ Theor.\ Phys.\ Suppl.\ }{\bf #1}}
\newcommand{\AP}[1]{{\it Ann.\ Phys.\ }{\bf #1}}


\begin{thebibliography}{100}

\bibitem{wz} 
J. Wess and B. Zumino, 
{\it Phys. Lett.} {\bf B49}, 52(1974). 

\bibitem{va} 
D.V. Volkov and V.P. Akulov, 
{\it Phys. Lett.} {\bf B46}, 109(1973). 

\bibitem{DZ} 
S. Deser and B. Zumino, 
{\it Phys. Rev. Lett.} {\bf38}, 1433(1977). 

\bibitem{ks1} 
K. Shima, 
{\it Z. Phys.} {\bf C18}, 25 (1983); \\
K. Shima, 
{\it European. Phys. J.} {\bf C7}, 341(1999). 

\bibitem{ks3} 
K. Shima, hep-ph/0012320, Phys. Lett. B in press. 

\bibitem{st1} 
K. Shima and M. Tsuda, hep-th/0012235. 

\bibitem{rikuzw} 
N.S. Baaklini, 
{\it Phys. Lett.} {\bf 67B}, 335(1976). 

%
\end{thebibliography}
\end{document}